\documentclass[pdflatex,sn-mathphys-num]{sn-jnl}

\usepackage{graphicx}%
\usepackage{multirow}%
\usepackage{amsmath,amssymb,amsfonts}%
\usepackage{amsthm}%
\usepackage{mathrsfs}%
\usepackage[title]{appendix}%
\usepackage{xcolor}%
\usepackage{textcomp}%
\usepackage{manyfoot}%
\usepackage{booktabs}%
\usepackage{algorithm}%
\usepackage{algorithmicx}%
\usepackage{algpseudocode}%
\usepackage{listings}%
\usepackage{soul} 

\theoremstyle{thmstyleone}%

\theoremstyle{thmstyletwo}%

\theoremstyle{thmstylethree}%

\begin{document}

\title[Article Title]{A critical consideration of X-ray detectors based on Ga$_2$O$_3$: excitation, carrier transport mechanisms and performance standardization}

\author[1]{\fnm{Alfred} \sur{Moore}}\email{alfred.moore@swansea.ac.uk}\equalcont{These authors contributed equally to this work.}

\author[2]{\fnm{Daniel A} \sur{Lamb}}\email{d.a.lamb@swansea.ac.uk}
\equalcont{These authors contributed equally to this work.}

\author*[1,2]{\fnm{Lijie} \sur{Li}}\email{l.li@swansea.ac.uk}
\equalcont{These authors contributed equally to this work.}

\author[3]{\fnm{Oliver} \sur{Fox}}\email{oliver.fox@diamond.ac.uk}

\author[3]{\fnm{Kawal} \sur{Sawhney}}\email{kawal.sawhney@diamond.ac.uk}

\author[2]{\fnm{Ciaran } \sur{Llewelyn}}\email{c.p.llewelyn@swansea.ac.uk}

\author[2]{\fnm{Jon E} \sur{Evans}}\email{j.e.evans@swansea.ac.uk}

\author[1]{\fnm{Saqib} \sur{Rafique}}\email{saqib.rafique@swansea.ac.uk}

\author[4]{\fnm{Tiantian} \sur{Chai}}\email{whxb4572@leeds.ac.uk}

\author[4]{\fnm{John} \sur{Harrington}}\email{J.P.Harrington@leeds.ac.uk}

\author[4]{\fnm{Zabeada} \sur{Aslam }}\email{Z.P.Aslam@leeds.ac.uk}

\author[4]{\fnm{Andrew P} \sur{Brown }}\email{A.P.Brown@leeds.ac.uk}

\author[4]{\fnm{Rik} \sur{Drummond-Brydson  }}\email{R.M.Drummond-Brydson@leeds.ac.uk}

\author*[1,2]{\fnm{Yaonan}\sur{Hou}}\email{yaonan.hou@swansea.ac.uk}

\affil*[1]{\orgdiv{Department of Electric and Electrical Engineering}, \orgname{Swansea University}, \orgaddress{\street{Bay Campus}, \city{Swansea}, \postcode{SA1 8EN}, \country{UK}}}

\affil*[2]{\orgdiv{Centre for Integrative Semiconductor Materials (CISM)}, \orgname{Swansea University}, \orgaddress{\street{Bay Campus}, \city{Swansea}, \postcode{SA1 8EN}, \country{UK}}}

\affil[3]{\orgdiv{Diamond Light Source}, \orgname{Harwell Science and Innovation Campus}, \city{Didcot}, \state{Oxfordshire}, \postcode{OX11 0DE},  \country{UK}}

\affil[4]{\orgdiv{School of Chemical and Process Engineering and the Bragg Centre for Materials Research}, \orgname{University of Leeds}, \city{Leeds}, \postcode{LS2 9JT},  \country{UK}}

\abstract{ X-ray detection underpins a wide range of applications in medicine, security, industrial inspection, scientific research for non-destructive imaging and material analysis. The rapid development of Ga$_2$O$_3$-based X-ray detectors offers a promising pathway toward next-generation detectors with high sensitivity, low noise, and harsh environment applications, benefiting from its intrinsic material properties such as high density, wide band gap energy, and high thermal-chemical stability. However, the underlying device operating mechanisms, including both carrier excitation and transport processes, have not yet been adequately studied, largely due to the misuse of X-ray sources in previous studies. Besides, benchmarking of device characteristics has been problematic due to experimental or data analysis issues, as well as misunderstandings of the applied equations associated with parameter definitions. In this work, we have designed and performed an instructive research work based on epitaxial $\beta-$Ga$_2$O$_3$:Si and its planar Schottky detectors, measured with energy-tunable monochromatic X-ray beams on a synchrotron beamline, clarifying the device excitation and carrier transport mechanisms with properly benchmarked device performance. In the end, we propose a set of protocols for correctly measuring and analysing the device performance. The proposed protocols are broadly applicable and can be readily extended to other semiconductor X-ray detectors.}


\maketitle

\section{Introduction}\label{sec1}
X-ray radiation constitutes a significant part of the electromagnetic spectrum, with a wavelength range from $\sim$10 $nm$ to $\sim$10 $pm$ (as shown in $\bold{Fig.1a}$). The high photon energy enables a light-matter interaction distinct from that of other wavelengths (e.g., UV-visible-infrared light), which leverages a broad applications in life science, semiconductor and steel industry, food and agriculture, healthcare and medical radiography, safety screening and astronomy.\cite{rawson2020x,wilson1947x,zschech2008high, mathanker2013x,spahn2013x, connolly2008x, wilkes2022x} As a result, efficient and reliable X-ray detection is highly demanded in advancing both scientific research and practical applications. 

The fast-developing semiconductor X-ray detectors can directly convert the radiant light into a photocurrent signal through the photoelectric effect, offering better conversion efficiency, spatial resolution, and response speed over the indirect detection scheme involving the utilization of scintillators. So far, the commercially available semiconductor X-ray detectors have mainly been fabricated from Si, primarily because of the mature complementary metal-oxide-semiconductor (CMOS) technology. However, such detectors suffer from several limitations, including low attenuation coefficient at relatively higher energy (e.g., $>$10 keV) due to the low atomic number (Z=14) /density (2.33 $g/cm^3$); high-voltage supply and slow response due to bulk material requirement; high thermal and ambient light noise; and low radiation hardness. Recent advances in Ga$_2$O$_3$ provide a promising solution for developing next-generation X-ray detectors with high absorption, low noise and harsh environment applications, benefiting from material advantages including a high density of 6.44 $g/cm^3$, a wide band gap energy of $\sim$4.95 eV, high-power compatibility, high thermal-chemical and radiation stability.

In recent years, various Ga$_2$O$_3$-based X-ray detectors have been developed, including vertical metal-semiconductor-metal (MSM) architectures on bulk materials, thin-film Schottky MSM structures, heterojunction diodes, transistor-like dose readers, flexible detectors, nanostructured sensors, and imaging sensors.\cite{lu2018schottky,langpoklakpam2025state,zhang2021varepsilon,chen2021high,shao2025amorphous,prasad2023ga2o3,kim2025high,zhang2023x,lu2019x} These detectors feature high sensitivity (up to 3.72 $\times 10^5 \mu C\cdot Gy_{air}^{-1}\cdot cm^{-2}$),\cite{yu2025high,gan2024sensitive} large photo-to-dark current contrast ratio (e.g., $\gg10^2$),\cite{chen2021fast}
fast response speed for pulsed radiant measurements (e.g., $<$20 ms),\cite{zhou2021pulsed} and high spatial resolution (e.g., $\sim$200 $\mu m^2$).\cite{liang2024retina} Nevertheless, most of the reported Ga$_2$O$_3$-based X-ray detectors, if not all, are characterized by using an X-ray tube, which is a broadband light source comprising both characteristic lines of the target material and broad Bremsstrahlung background (containing both soft and hard X-ray radiation). \cite{ebel1999x} It is worth noting that soft X-rays overlap with deep ultraviolet (DUV) light in the low-energy region on the electromagnetic spectrum ($\bold{Fig.1a}$), where the distinction between the two depends primarily on their generation mechanisms. Consequently, the photocarrier generation in Ga$_2$O$_3$ detectors under such illumination could involve two major coexisting mechanisms: the near-band edge absorption at the lower-energy region (as shown in $\bold{Fig.1b}$) and core electron transition at the higher-energy region (as shown in $\bold{Fig.1c}$). As a result, the carrier excitation mechanism (and potentially carrier transport) could not be adequately resolved. Besides, the benchmarking of device characteristics has been found problematic due to either the experimental setup or the misunderstanding of the equations/parameters in data analysis. 

\begin{figure}[h]
\centering
\includegraphics[width=1\textwidth]{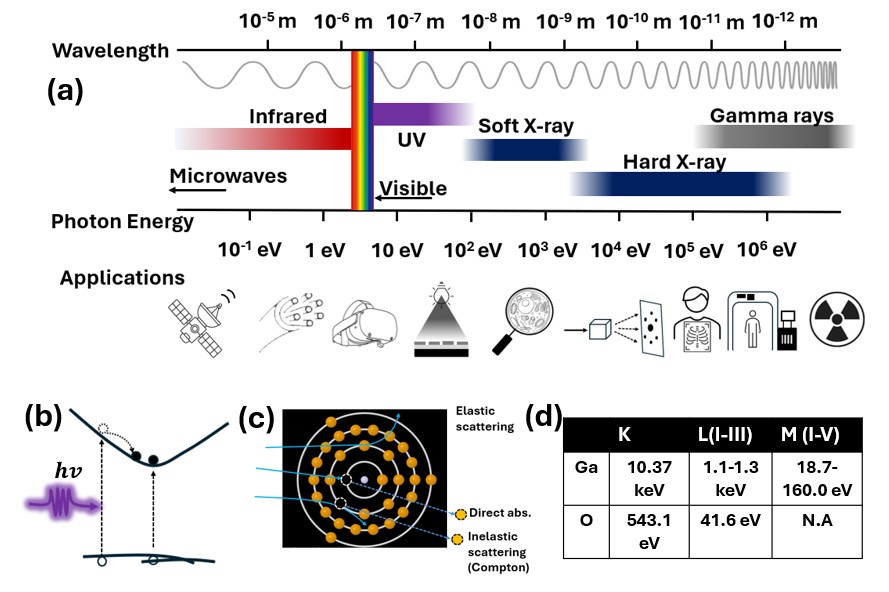}
\caption{ (a) X-ray waveband on the electromagnetic spectra and their application examples; (b) Schematic of near-band edge absorption of low energy photons (DUV and part of soft X-ray) in $\beta$-Ga$_2$O$_3$, where the Auger relaxation of an energized electron to conduction band minimum is also shown. (c) Schematic of X-ray interaction with Ga atoms, where the direct absorption and inelastic (Compton) scattering contribute to the internal photoelectric effects and elastic scattering does not generate energized electrons.\cite{markwinder2026} (d) Table of the electron binding energies of Ga and O atoms.\cite{bearden1967reevaluation,cardona1978photoemission,gwyn2000}}\label{fig1}
\end{figure}

To provide guidance for the further development of Ga$_2$O$_3$-based X-ray detectors, we have studied the excitation and carrier transport based on a planar Schottky MSM device fabricated from $\beta-$Ga$_2$O$_3$ (the most stable phase among the 6 polymorphs) under \textit{tunable monolithic X-ray} radiation. The X-ray energy range from 8 to 24 keV was selected, which straddles the highest electron binding energy of 10.37 keV (Ga 1s electrons) for deciphering the excitation mechanism, whilst the carrier transport process was elucidated by analyzing the current-voltage (IV) curves with different Si doping concentrations. At the end of our instructive research, we have proposed a series of protocols for standardized measurements of key device parameters and the correct utilization of mathematical equations to benchmark the performance of Ga$_2$O$_3$-based X-ray detectors.

\section{Results}\label{sec2}

\subsection{Samples and devices}\label{subsec2.1}
The $\beta-$Ga$_2$O$_3$ with a thickness of $\sim$2.03 $\mu m$ (inset of $\bold{Fig.2a}$) were grown on 0.2$^\circ$ off-cut C-plane sapphire substrate with Si doping concentration ranging from $\sim$10$^{15} \, atoms\, \cdot  cm^{-3}$ to $\sim$10$^{20} \, atoms\, \cdot cm^{-3}$. The devices fabricated with 6.3$\times$ $10^{18} \, atoms\, \cdot cm^{-3}$ (labelled as Sample/Device $\bold {A}$) and 5.2$\times$ $10^{15} \, atoms\, \cdot cm^{-3}$ (labelled as Sample/Device $\bold {B}$) have been selected for detailed study considering the evolution of the photocurrent under X-ray radiation. The Si doping concentration is determined by secondary-ion mass spectrometry (SIMS) measurements ($\bold{Fig.S1}$, supplementary information). $\bold{Fig.2a}$ shows the X-ray diffraction (XRD) in $\theta-2\theta$ scan mode, where single crystalline with (-201) epitaxial plane was observed. The optical band gap is 4.9 eV confirmed by the transmittance measurement ($\bold{Fig.2b}$). X-ray photoelectron spectroscopy (XPS) exhibits an asymmetry O-1s peak, which is further decomposed into two peaks at $\sim$ 527.5 eV and $\sim$ 529 eV ($\bold{Fig.2c}$). The former is usually attributed to the Ga-O bonding, while the higher energy component is associated with oxygen vacancy (V$_o$).\cite{liang2024retina} Obviously, the density of V$_o$ is reduced with increasing Si doping, which is believed to be a result of the formation of defect complexes.\cite{shokri2023point} Ti (20 nm)/Au (100 nm) interdigital electrodes with a length of 1000 $\mu m$ and spacing of 50 $\mu m$ have been fabricated on both samples to form the MSM diodes as shown in $\bold{Fig.2e}$. $\bold{Fig.2d}$ shows measurement setup using \textit{tunable monolithic X-ray} generated from a synchrotron light source, where the device under measurement (DUT) is mounted on an automated translation stage with the same plane of a dosemeter before an alignment camera. $\bold{Fig.2f}$ shows the photo of the device (bonded on a printed circuit board) under an 8 keV X-ray beam. It is worth to note that the whole device including the electrodes is transparent under the minimum energy (8 keV) used in our experiments. 

\begin{figure}[h]
\centering
\includegraphics[width=1\textwidth]{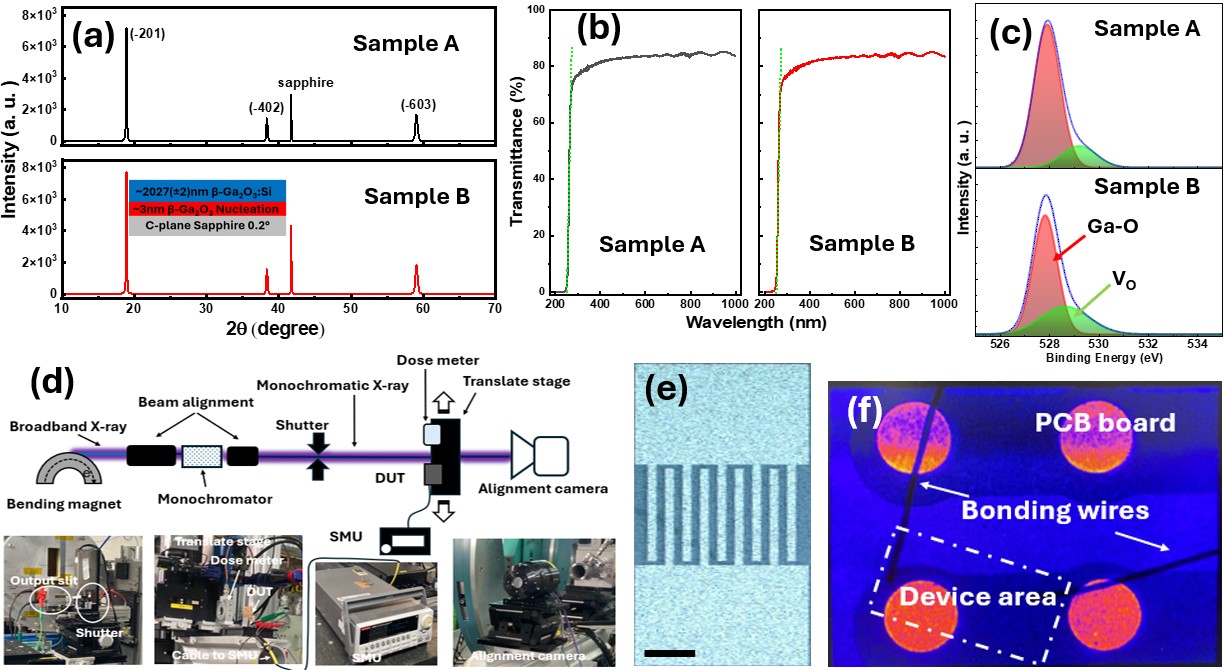}
\caption{ (a) XRD $2\theta$ scan of the epitaxial films, which confirms single $\beta$-phase Ga$_2$O$_3$ formed; (b) Transmittance spectra show both samples a sharp absorption at 253 nm (4.9 eV). (c) XPS spectrum of both samples. High Si-doped sample exhibits a substantial reduction in the V$_o$ peak. (d) Top: Schematic of the measurement setup; Bottom: photos of some key components of the measurement setup. (e) Optical image of the device (scale bar = 500 $\mu m$); (f) Photo of the DUT bonded on PCB from the alignment camera, where the entire device is transparent to X-ray (with an energy of 8 keV).}\label{fig2}
\end{figure}

\subsection{Photocurrent and carrier transport}\label{subsec2.2}

The current-voltage scans under dark and X-ray radiation (12 keV radiation with a flux of $\sim 10^{9}$ cps) of all the devices with Si doping concentrations from $10^{15}$ $ \, atoms\, \cdot cm^{-3}$ to $10^{20}$ $\, atoms\, \cdot cm^{-3}$ are shown in $\bold{Fig.S2}$, with the statistical photocurrent values and contrast ratio shown in  $\bold{Fig.3a}$. Here, the photocurrent is defined as, 

\begin{equation}
I_{ph}=I_{tot}-I_{dark}.\label{eq1}
\end{equation}

\noindent where $I_{ph}$,$I_{tot}$, and $I_{dark}$ are photocurrent, total current under X-ray radiation, and dark current, respectively. With a high Si doping concentration $>6\times 10^{19} \space \, atoms\, \cdot cm^{-3}$, the device exhibits an Ohmic behavior under both dark and X-ray illumination. Although the $I_{ph}$ of the Ohmic devices has reached a value of $10$ mA, the contrast ratio falls to $<1$, unsuitable for practical applications. Therefore, we focus on the devices with Si doping concentrations $\leq 6.29 \times 10^{18} \space \, atoms\, \cdot cm^{-3}$. In this doping range, the photocurrent is monotonically increasing with the doping concentration while maintaining a reasonably good contrast ratio ($>10$), measured from the current-voltage scan. Focusing on Device $\bold{A}$ and $\bold{B}$, their looped dark IV curves are shown in $\bold{Fig.3b}$, exhibiting a dark current at $\leq 4 \times 10^{-9} A$ under 15 V. Both devices exhibit capacitive impedance in the low-voltage range ($-4$ to $4$ V), accompanied by current rippling, which is supposed to be related with the intrinsic defects.\cite{moore2025capacitance} The coupled capacitance decreases with the increasing conductivity of the material ($\bold{Fig.S3}$), and vanishes with large photocurrent (Device $\bold{A}$ in $\bold{Fig.3c}$), owing to the photogenerated carriers.

\begin{figure}[h]
\centering
\includegraphics[width=1\textwidth]{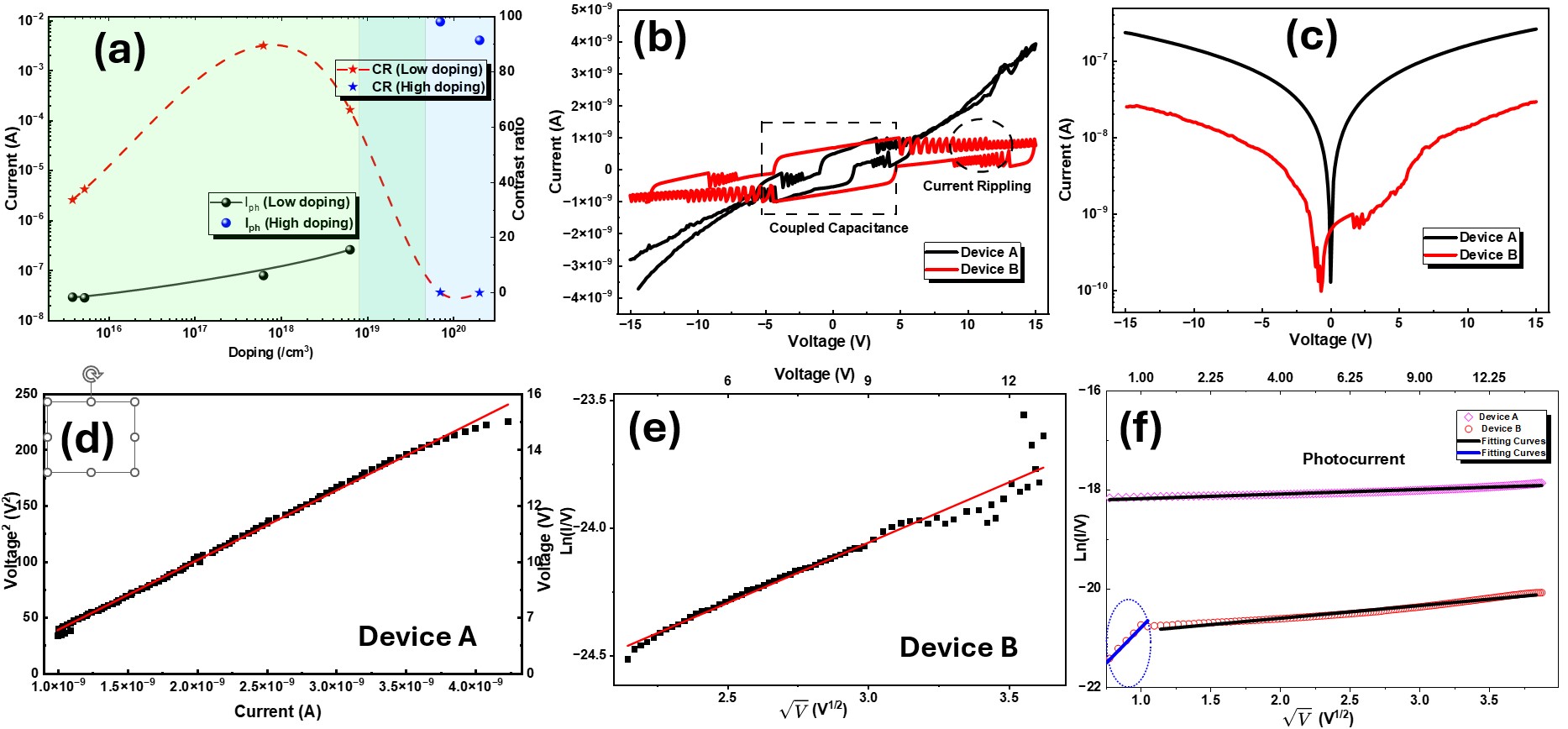}
\caption{ (a) Photocurrent and contrast ratio of the MSM devices fabricated on $\beta$-Ga$_2$O$_3$ with Si doping concentration from $10^{15}$ to $10^{21} \space \, atoms\, \cdot cm^{-3}$; (b) Dark current; and (c) photocurrent of Device $\bold {A}$ and $\bold {B}$; (d) Dark current of Device $\bold {A}$ plotted in $V^2$-I. (e) Dark current of Device $\bold {B}$ plotted in $ln (I/V)$-$\sqrt{V}$. The voltages in (c) and (d) are larger than 4 V to avoid the coupled capacitance influences in both devices. (f) Photocurrent of both devices plotted in $ln (I/V)$-$\sqrt{V}$. Note: the current of Device $\bold{B}$ under 2 V (marked area) is not taken into consideration as the coupled capacitance still exists (observed from Fig.3c)}\label{fig3}
\end{figure}

The dark current ripping reduces with high doping concentration, indicating the compensation of the intrinsic defects. This is further confirmed by analyzing the carrier transport. With a doping of 6.29  $\times 10^{18} \space \, atoms\, \cdot cm^{-3}$ (Device $\bold {A}$), the IV can be fitted into a linear relationship following Mott's space-charge-limited current (SCLC) theory, ($\bold{Fig.3d}$),\cite{mott1948electronic}

\begin{equation}
I_{d}=\frac{9A\mu \epsilon V^2}{8l^3}.\label{eq2}
\end{equation}

\noindent where $A$ is the contact area, $\mu$ the field dependent mobility, $\epsilon$ the dielectric constant and $l$ the distance of the electrodes. The SCLC is merely related to injected carriers and not relevant to the defects. Therefore, it is often reasonably attributed to charge-filled traps, \cite{carbone2005space,zhang2025effect} in our case, resulting from Si doping. In contrast, the IV relationship of the low-doping device ($\bold {B}$) can only be fitted into charge-trap related Poole-Frenkel current by, \cite{sze2021physics,rafiq2018carrier}

\begin{equation}
I_{d} \propto V exp(\frac{\beta_{PF} \sqrt{V/d}}{kT}).\label{eq3}
\end{equation}

\noindent where $k$ is the Boltzmann constant, and $\beta_{PF}$ the Poole Frenkel barrier reduced factor, which is a constant given by $\beta_{PF}=\sqrt{q^3/\epsilon_0 \epsilon_r}$ ($q$, $\epsilon_0$ and $\epsilon_r$ are unit charge, vacuum permittivity and relative permittivity, respectively). As shown in $\bold{Fig.3e}$, a good linear fitting between $\ln(I/V)$ and $\sqrt{V}$ (modification of Eq. \ref{eq3}) indicates the existence of defects, which act as charge traps. The reason is that the intrinsic defects, which can contribute to a background carrier concentration up to $10^{18} \, cm^{-3}$ in an undoped sample (typically $\sim 10^{16} \, cm^{-3}$ from the unintentionally doped sample grown by our team),\cite{lorenz1967some} cannot be sufficiently passivated with a low Si doping concentration ($5.24 \times 10^{15} \space \, atoms\, \cdot cm^{-3}$ for Device $\bold {B}$). This conclusion also corresponds well to the XPS observations in $\bold{Fig.2c}$. Under X-ray radiation, the photocurrent of all devices follows the Poole-Frenkel current as shown in $\bold{Fig.3f}$ (devices with other doping shown in $\bold{Fig.S4}$), due to the excitation of filled traps. 

\subsection{Transient response}\label{subsec2.3}

The transient response at various working voltages has been studied for a deeper insight of the device working mechanism. $\bold{Fig.4a}$ displays the photocurrent rise and decay of 
Device $\bold{A}$ with a relative high doping. The photocurrent decay can be fitted by a multi-term exponential decreasing function, \cite{hou2025photocurrent,ouyang2025performance}

\begin{equation}
I_{t} = I_0 + \sum_{k=1}^{n} A_n exp(-\frac{t-t_0}{\tau_n}).\label{eq4}
\end{equation}

\noindent where the $I_0$, $A_n$ and $t_0$ are fitting parameters, and $\tau_0$ the decay constant related to the carrier relaxation process. Similarly, the photocurrent rise can be fitted by multiple exponential growth (e.g, decay to saturation),

\begin{equation}
I_{t} = I_0 - \sum_{k=1}^{n} A_n exp(-\frac{t-t_0}{\tau_n}).\label{eq5}
\end{equation}

\noindent The number $n$ denotes the number of physics processes identified that are associated with carrier rise or decay. The photocurrent decay of Device $\bold{A}$ can be well fitted into a 3-term exponential decay function, with time constants (in $\bold{Table \, 1}$) very similar to those of the MSM device fabricated on an amorphous Ga$_2$O$_3$ in our previous report.\cite{hou2025photocurrent} Therefore, the fast ($\tau_1$), medium ($\tau_2$) and slow ($\tau_3$) decays of Device $\bold{A}$ are attributed to defect-midiated carrier recombination, carrier de-trapping, and slow motion related to the deep-level defects, such as $V_o$ and its charged states ($V_o^*$).\cite{hou2025photocurrent} The result also indicates that the photo-generated carrier transport is independent of the excitation mechanism (e.g., either by deep ultraviolet or X-ray). The rise time of both devices is fitted into a 2-term exponential growth, suggesting two observable excitation processes. As the photo-excitation is an ultrafast process, \cite{koksal2018measurement} the observed rise time constants on the second and 10-second level are likely due to establishing the balance of trapping/de-trapping associated with two kinds of deep-level defects in the material. 

\begin{figure}[h]
\centering
\includegraphics[width=1\textwidth]{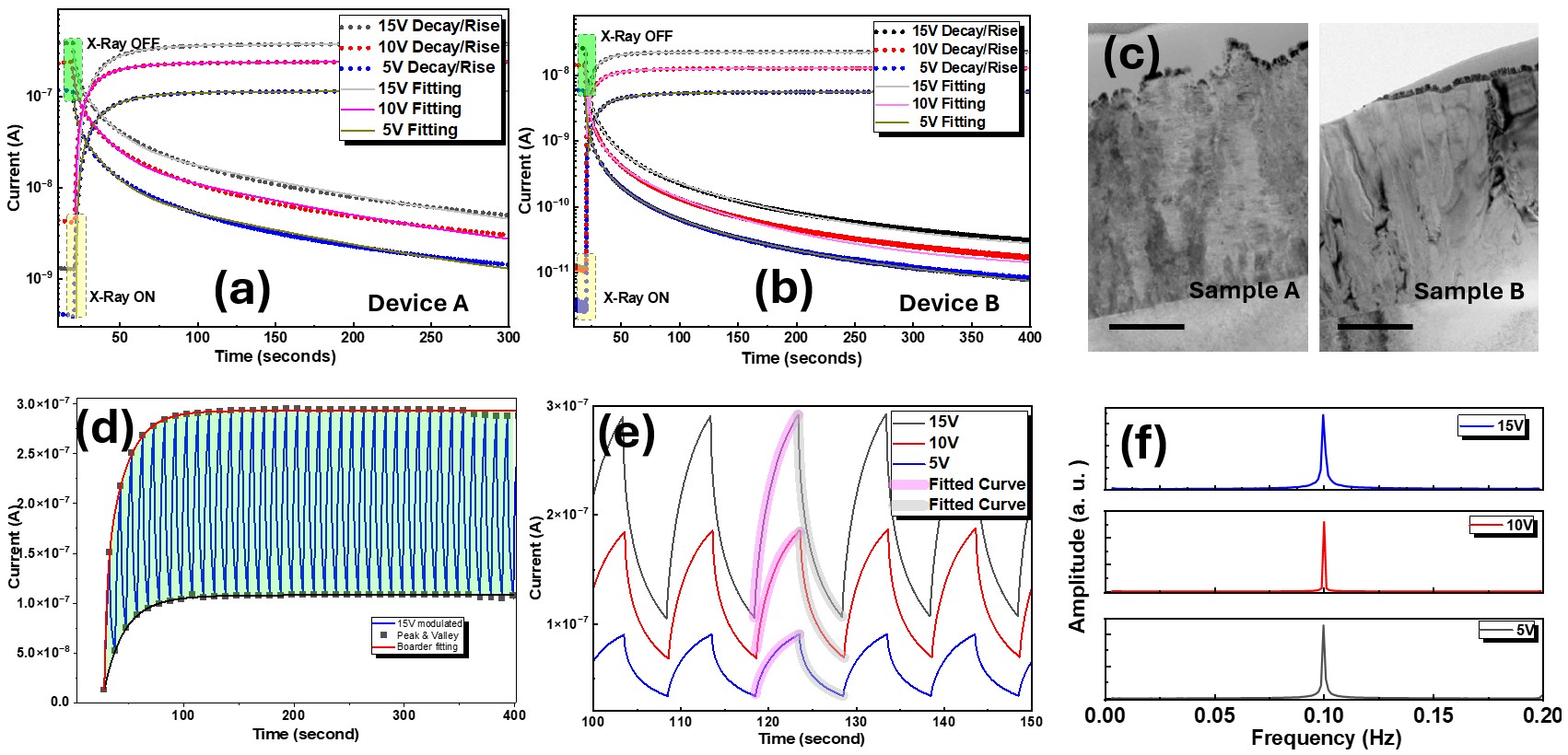}
\caption{ Photocurrent rise and decay at different voltages (5, 10, 15V) of Device \textbf{A} (a) and \textbf{B} (b); (c) Bright field TEM images of cross-section of Samples with high ($\times10^{20} \, atoms \,cm^{-3}$, left) and low ($\times10^{15} \, atoms \,cm^{-3}$, right) Si concentrations with similar growth conditions, scale bar = 500 nm; (d) Pulsed photocurrent under modulated X-ray radiation on Device \textbf{A}; (e) Zoom-in view of the pulsed photocurrent in the range of 100-150s. (f) Frequency of the modulated X-ray by Fourier Transform of the pulsed photocurrent at different voltages.}\label{fig4}
\end{figure}

The photocurrent decay with a low-doping concentration (Device \textbf{B}) can on longer been fitted with Eq. \ref{eq4}. Instead, a multi-term stretched exponential decay have to be used (detailed discussion in $\bold{Fig.S5}$),

\begin{equation}
I_{t} = I_0 + \sum_{k=1}^{n} A_n exp(-(\frac{t-t_0}{\tau_n})^{\beta_{n}}).\label{eq6}
\end{equation}

\noindent where the stretch factor $\beta_n$ is related to the spatial fluctuations of the carrier traps. In other words, the photocurrent decay is not only related to the type of defects, but also to their distribution, akin to the carrier recombination process in temporal photoluminescence observations. \cite{lee2001application} Moreover, this decay process is strongly related with the applied voltage. As can be seen from the fitting results in $\bold{Table \, 1}$, a 2-term stretched-exponential decay function can satisfy the photocurrent decay trace at low voltage (5 V), but a 3-term function is required for higher voltages. Compared with the applied voltages of 10 and 15 V, the decay constants (e.g., $\tau_1$ and $\tau_2$) are similar, but with significantly different stretch factors, which can be attributed to the inhomogeneous distribution of the impurities. This is further verified by the Transmission Electron Microscope (TEM) observations.  As illustrated in \textbf{Fig.4c}, the sample with low doping exhibits large structural defects randomly distributed in both horizontal and vertical directions, whereas higher Si doping significantly improves the uniformity of the crystal grains. In the meantime, the energy dispersive spectroscopy (EDS) reveals a more uniform Si distribution in the material (\textbf{Fig.S6}). Those non-uniformly distributed defects interact with the applied electric field, whose depth is governed by the applied voltage,\cite{hou2014monolithic} leading to the voltage-dependent stretch factors. 

From the \textit{saturated photocurrent} shown in \textbf{Figs. 4a} and \textbf{4b}, the sensitivity can be calculated by,\cite{pan2017cs2agbibr6,yakunin2016detection} 

\begin{equation}
S = \frac{I_{ph}}{\dot{D}_{air}A_{op}}=\frac{I_{tot}-I_{dark}}{\dot{D}_{air}A_{op}}.\label{eq7}
\end{equation}

\noindent where $\dot{D}_{air}$ and $A_{op}$ are the dose rate normalized to air, \cite{sakhatskyi2025characterizing} and optical active area of the device. Following Eq \ref{eq7}, the sensitivity of Devices \textbf{A} and \textbf{B} are $S_A(12 keV)=1.42\, \mu C\cdot mGy \cdot cm^{-2}$, and $S_B(12 keV)=0.09\, \mu C\cdot mGy \cdot cm^{-2}$ under 15 V, with a $I_{ph}/I_{dark}$ contrast ratio over 3 orders of magnitude. The corresponding bulk sensitivities are $S_A(12 keV)=7.03\times10^3\, \mu C\cdot mGy \cdot cm^{-3}$ and $S_B(12 keV)=446\, \mu C\cdot mGy \cdot cm^{-3}$ by considering the material thickness.

\begin{table}[h]
\caption{Decay and rise constants obtained from fitting using Eq. (4)-(6)}\label{tab2}
\begin{tabular*}{\textwidth}{@{\extracolsep\fill}lcccccc}
\toprule%
& \multicolumn{3}{@{}c@{}}{Device A} & \multicolumn{3}{@{}c@{}}{Device B} \\\cmidrule{2-4}\cmidrule{5-7}%
Condition & $\tau_1 (s)$ & $\tau_{2} (s)$ & $\tau_{3} (s)$ & $\tau_1 (s)$ /$\beta_1$ & $\tau_{2} (s)$ /$\beta_2$ & $\tau_{3} (s)$/$\beta_3$ \\
\midrule
Decay (15 V)  & 1.78 & 14.00 & 114.06 & 0.36/0.64 & 1.98/0.56 & 7.89/0.44 \\
Decay (10 V)  & 1.82 & 14.12  & 109.85  & 0.32/0.86 & 1.70/0.62 & 6.55/0.46\\
Decay (5 V)\footnotemark[1]  & 1.85 & 14.06 & 108.38 & 0.65/0.38 & 3.87/0.41 & -/-\\
Rise (15 V)  & 18.44 & 82.79 & - & 6.20 & 18.79 & - \\
Rise (10 V)  & 18.56  & 85.60 & - & 5.25 & 21.36  & - \\
Rise (5 V)  & 18.96 & 85.89 & - & 11.26 & 36.75 & - \\
\botrule
\end{tabular*}

\footnotetext[1]{2-term stretched exponential decay satisfies the fitting of decay trace of Device $\bold{B}$ under 5 V.}

\end{table}

Although long decay time finds its application in the emerging optical synapse,\cite{liang2024retina} a fast response speed is generally preferred for the pulsed/modulated X-ray detections.\cite{zhou2021pulsed,chen2021fast,chen2021highly} Regarding this, the response of our devices under modulated X-ray radiation has been examined under various applied voltages. As a representative, $\bold{Fig. 4d}$ exhibits the photocurrent rise of Device $\bold{A}$ until saturation, under a square wave modulated X-ray with a period of 10s and a duty cycle of 0.5 (a full data set for both devices can be found in $\bold{Figs.S7}$ and $\bold{S8}$, supplementary information). The stable peak current is found to be similar to the saturated current obtained in \textbf{Figs.4a} and \textbf{4b} with the same dose rate, though it does not return to the dark current. Therefore, the device could be used to evaluate the incident dose of the modulated X-ray beam. The transient photocurrent, including the peak and valley photocurrent evolution in $\bold{Fig. 4d}$,  pulsed photocurrent growth and decay in $\bold{Fig. 4e}$, can be fitted with Eq. \ref{eq4} and \ref{eq5}. However, the obtained time constants ($\bold{Table \, S1}$) are different to those obtained in $\bold{Table \, 1}$ considered enough device response time. Therefore, the device response speed cannot be accurately evaluated using pulsed X-ray radiation when the modulation rate exceeds the intrinsic response speed; this limitation has been overlooked in some previous studies (see detailed discussion in later sections). \cite{gan2024sensitive, lu2019x} 

To measure the modulated X-ray properties, we propose a modulated depth factor of the detector, 

\begin{equation}
MD = \frac{I_{peak}-I_{valley}}{I_{ph}} = \frac{I_{peak}-I_{valley}}{I_{tot}-I_{dark}} .\label{eq8}
\end{equation}

\noindent where the $I_{peak}$ and $I_{valley}$ are the saturated peak and valley photocurrent shown in $\bold{Fig. 4c}$. This factor is irrelevant to the applied voltage, as statically shown in $\bold {Table \, 2}$, and thus only related to the modulation frequency ($f$), which can be obtained by Fourier Transform of the pulsed photocurrent (\textbf{Fig.4f}). Once the MD is characterized with $f$, the sensitivity under modulated X-ray radiation can be further rewritten as $S_{pulse}=MD_{f}S$. Together with Eqs \ref{eq7} and \ref{eq8}, the dose rate of the incident pulsed X-ray can be measured once $I_{peak}$ and $I_{valley}$ are obtained. This process also avoids any potential issue of $\frac{I_{peak}}{I_{tot}}\not= \frac{I_{valley}}{I_{dark}}$ by only assuming the $I_{peak}\simeq I_{tot}$. 

\begin{table}[h]
\caption{MD values obtained from the modulated X-ray radiation}\label{tab2}%
\begin{tabular}{@{}llll@{}}
\toprule
Device\footnotemark[1] & MD (15 V)  & MD (10 V)  & MD (5 V) \\
\midrule
Device $\bold{A}$    & 0.85   & 0.86  & 0.86  \\
Device $\bold{B}$    & 0.49   & 0.50  & 0.49  \\
\botrule
\end{tabular}

\footnotetext[1]{The small variation in the second decimal place is from number rounding.}

\end{table}

\begin{figure}[h]
\centering
\includegraphics[width=1\textwidth]{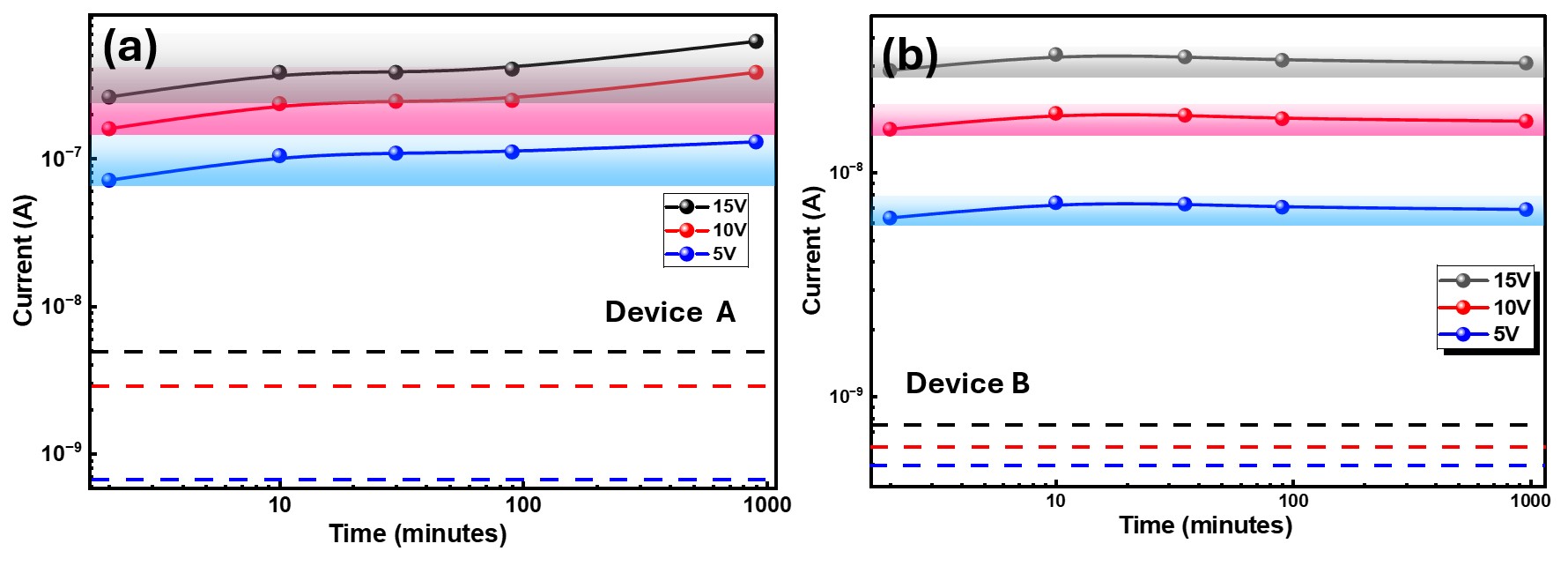}
\caption{ Photocurrent stablity of (a) Device $\bold {A}$,  (b) Device $\bold {B}$. The X-ray energy and dose rate are 12 keV and $10^{10}$ cps, respectively. }\label{fig5}
\end{figure}

Although the thermochemical stability and radiation hardness of Ga$_2$O$_3$ have been frequently reported in the literature, its potential as an X-ray detector has not been explored. To do this, we measured the photocurrent under 15 V under high flux radiation ($1.7/.8\times 10^{9} \, cps\cdot cm^{-2}$ at 12 keV) up to 17 hours. As shown in $\bold{Fig. 5a}$, instead of degradation, the photocurrent exhibits $>2$-fold improvement in Device $\bold{A}$, which is likely due to the Si dopants ionized (activated) by the long-term X-ray radiation.\cite{4395007} In contrast, the photocurrent of Device $\bold{B}$ (low doping) is fairly stable as shown in  $\bold{Fig. 5b}$, verifying the material and device radiation hardness. 

\section{Discussion}\label{sec3}

\subsection{Absorption and excitation}\label{subsec3.1}

$\bold{Fig.6a}$ shows measured linear attenuation coefficient ($\alpha$) of $\beta-$Ga$_2$O$_3$ in the range of 8-24 keV along the crystallographic direction of [-201] on a 3 $\mu m$ film, where $\alpha$ is calculated following the Beer–Lambert law, $I(x)=I_0e^{-\alpha x}$ ($I_0$ and $x$ and are the incident X-ray dose and the material thickness, respectively). As one can see, the measured data generally matches the theoretical value calculated through the NIST database, \cite{NIST}
while our measured data indicates a higher Ga 1s electron absorption ($>$1300$ \, cm^{-1}$) than the theoretical value ($\sim$1100$ \, cm^{-1}$).  This can be attributed to the theoretical calculation not taking into account the monoclinic polymorph, which likely underestimates the X-ray scattering in the single crystal. Therefore, we shared our measured data and fitting results to support the fast-growing research in Ga$_2$O$_3$-based X-ray detectors (Supplementary Information, \textbf{Table. S2}). 

As the linear attenuation coefficient reflects the X-ray absorbed by the material, it is rational to infer that the quantum efficiency (QE) of the detector should follow the same manner. As such, we performed the QE mapping of Device \textbf{A} (details of Device \textbf{B} can be found in the Supplementary Information, \textbf{Figs. S9} and \textbf{S10}) as a function of X-ray energy and the voltage (\textbf{Fig. 6b}), where the maximum value can be found at $\sim$10.5 keV. For a better visualisation, \textbf{Fig.6c} shows the measured QE under 15 V, which matches well with the fitting curve of the linear absorption coefficient obtained from \textbf{Fig.6a}. Therefore, it is concluded that the photocarrier generation is dominated by the direct absorption by core electrons near the binding energy (e.g., Ga 1s electrons at $\sim$10.5 keV), then shifts into a scattering dominant mechanism at other energies. In addition, the QE is much larger than 1, due to the defect-mediated charge carrier transport to be discussed later.

$\bold{Figs.6d}$ shows the sensitivity in the measured energy range, where a slight increase from 0.5 to 4 $\mu C\cdot mGy_{air}\cdot cm^{-2}$ due to the interplay of photon energy and the air mass absorption coefficient. $\bold{Figs.6e}$ shows the dose-dependent photocurrent mapping, where a good linearity can be found between photocurrent and dose. $\bold{6f}$ displays the photocurrent as a function of the product of dose rate and the optical active area under 5, 10 and 15 V, respectively, ranging from $\sim 10^{-3}$ to $>10^{-1}$ $mGy\cdot cm^2\cdot s^{-1}$ under 12 keV. A good linear fit of the data is observed, where the slope reflects the device sensitivity (e.g., 0.92 $\mu C\cdot mGy_{air}\cdot cm^{-2}$ under 15V). The sensitivity acquired by the mapping methods is smaller than that obtained from the $saturated \, photocurrent$ in  $\bold{Figs.4a}$, resulting from the tradeoff between scan speed and the device transient properties. Therefore, we propose a standardization of benchmarking the performance for Ga$_2$O$_3$-based X-ray detectors in the following section.

\begin{figure}[h]
\centering
\includegraphics[width=1\textwidth]{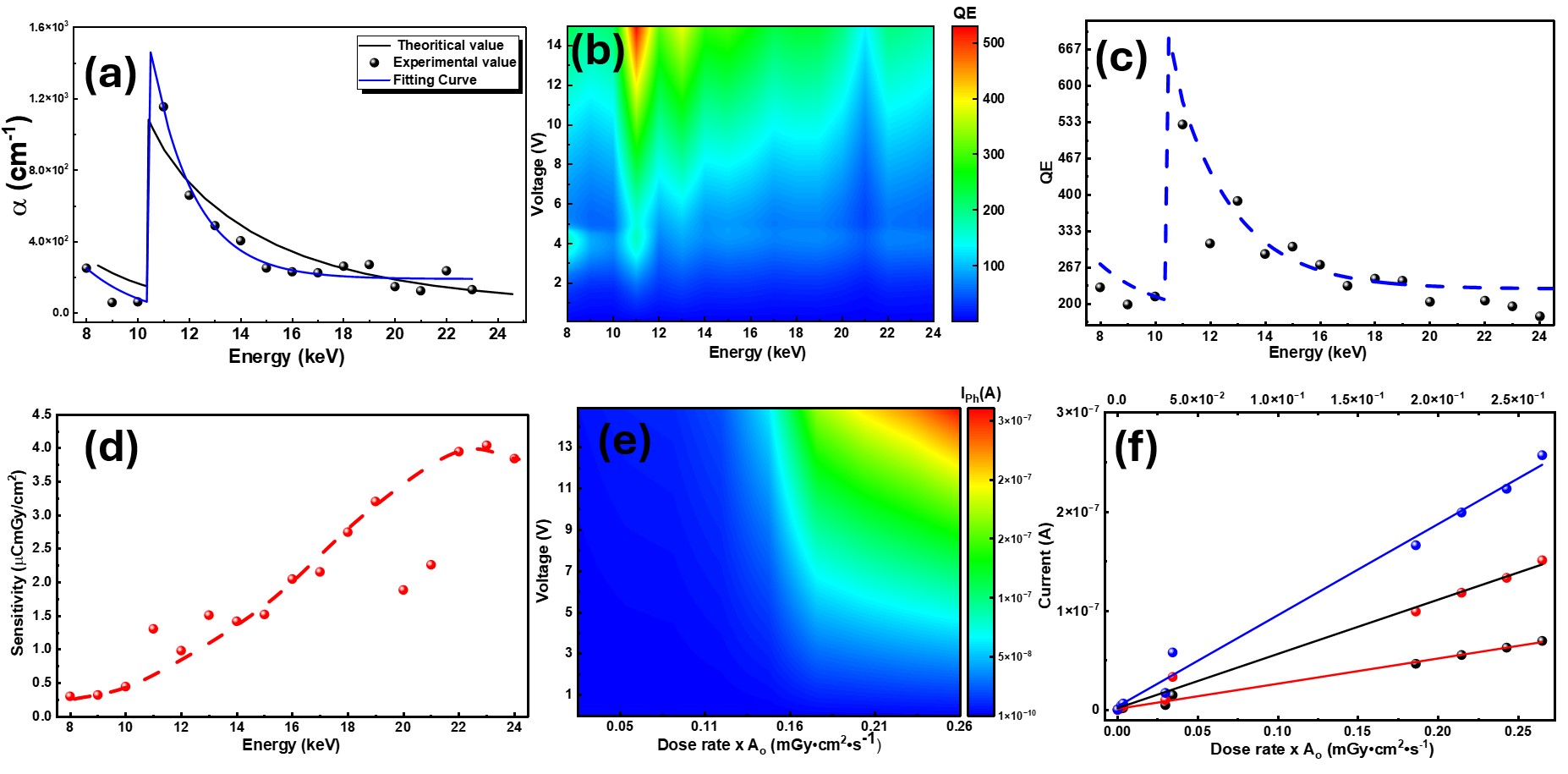}
\caption{ (a) Theoretical (dark), experimental (dots) and fitting curve (blue) of linear attenuation coefficient; (b) Scanned QE under different energies; (c) QE plot as a function of energy at 15 V, where the blue dashed curve is the fitting curve from \textbf{Fig.6a}; (d) Scanned sensitivity at 15 V under different energies; (e) Dose-dependant photocurrent mapping; (f) Photocurrent plotted as a function of the dose rate $\times A_o$.}\label{fig6}
\end{figure}

\subsection{Standardization of device benchmarking }\label{subsec3.1}

Sensitivity is a key characteristic of X-ray detector. An improperly measured or calculated value will not only hinder benchmarking of X-ray detectors but also lead to problems in deriving other related parameters (e.g., quantum efficiency, noise level, and limit of detection). Furthermore, the inappropriately benchmarked device parameters (e.g., sensitivity) prevent the comparison between reports, eventually damaging the development X-ray detectors based on this emerging material. Therefore, here we propose a protocol for correctly obtaining the sensitivity. Following Eq. \ref{eq7}, an accurate sensitivity requires precise measurement of $I_{ph}$ and $\dot{D}_{air}$, with a well-defined $A_{op}$ and following considerations,
\begin{itemize}
\item Sensitivity is a single-energy (or wavelength) specification, as the X-ray absorption in Ga$_2$O$_3$ is energy-dependent (as shown in $\bold {Fig. 6a}$). Therefore, it should always be stated with the excitation energy. From this point of view, the values measured using X-tube, which is subjected to the broad Bremstrahlung emission, are inaccurate (unless any extra spectral selection component has been utilized and reported). 
\item Dose rate of the monochromatic beam should be calibrated to air absorption instead of the material itself for a consistent calculation of the sensitivity, facilitating the comparison of devices. Some dosemeters read X-ray density in photon flux ($\Phi$) mode with an unit of $cps\cdot cm^{-2}$, which can be translated to $Gy\cdot s^{-1}$ by,

\begin{equation}
\dot{D}_{air} = (\frac{\mu}{\rho})_{air}E_{ph}\Phi.\label{eq9}
\end{equation}

\noindent where $(\frac{\mu}{\rho})_{air}$ is the mass photoabsorption coefficient of air, which is energy-dependent; $E_{ph}$ is the (monochromatic) X-ray photon energy. By examination of Eqs \ref{eq7} and \ref{eq9}, one can conclude that the sensitivity is nonlinear with the photon energy (as shown in \textbf{Fig. 6d}).
\item The measurement of $I_{ph}$ depends on the device properties and the experimental setup. It is worth mentioning that slow photocurrent rise and decay (from several seconds up to several hundred seconds) are often observed in Ga$_2$O$_3$ detectors due to the presence of deep-level defects or coupled capacitance (e.g. in this work). As such, photocurrent obtained from IV measurements, often with a relatively faster scan speed (e.g., 20 ms in this work), will not be accurate (although it is still an indicative comparison of devices fabricated from the same batch, such as $\bold{Fig.3a}$). It is recommended to use a \textit{saturated photocurrent} obtained in $I-t$ measurement ($\bold{Figs.4a}$ and $\bold{4b}$) with sufficient time to achieve a stable saturation. 
\item The slow response related with deep-level defects is often associated with a photocurrent gain through charge accumulation,\cite{hou2025photocurrent} defined as $G=\frac{I_{ph}}{I_{theo}}$, where $I_{theo}$ is the theoretical photocurrent supposing each X-ray photon generates an electron-hole pair. Therefore, the sensitivity calculated by Eq.\ref{eq7} will include such a gain. However, it is worth pointing out that the large sensitivity value due to defect-mediated carrier transport doesn't mean the detector is sensitive to low signals, which might not be enough to generate accumulated photogenerated carriers.\cite{gan2024sensitive} Instead, such a gain must be considered to calculate, for example, noise equivalent dose and detective quantum efficiency (DQE).\cite{sakhatskyi2025characterizing} 
\item Please be aware that $A_{op}$ is the optical active area (or effective area in some literature) in Eq. \ref{eq7}. This area is not the contact area, although they are occasionally the same (e.g., some vertical devices). When using the \textbf{Fig.6f} for obtaining the sensitivity, one must use the current plotted as a function of the product of $\dot{D}_{air} \times A_{op}$. The plot of current density (in unit of $mA\cdot cm^{-2}$) against $\dot{D}_{air}$ (in unit of $mGy\cdot s^{-1}$) is inaccurate (though the final unit is the same as that of sensitivity).\cite{chen2021fast,liang2024retina,sakhatskyi2025characterizing} 
\item The contact of Ga$_2$O$_3$-based devices is often made from a thin metal layer (e.g., $<200$ nm), which is transparent to hard X-rays (\textbf{Fig.2f}). It is suggested that the energy (range), contact stacks and thickness, and the defined $A_{op}$ be explicitly reported for properly benchmarking the sensitivity. For example, the interdigited MSM contact (broadly adopted in Ga$_2$O$_3$-based X-ray and DUV detectors) facilitates the charge carrier collection. But the whole area (including the contact area and the area between the interdigited fingers) is $A_{op}$, different from its DUV counterparts, whose  $A_{op}$ only considers the area between the interdigited fingers. Ingorning this can lead to a significantly overestimated sensitivity value.
\item The sensitivity calculated from Eq \ref{eq7} doesn't reflect the details of the device structure. Given that the linear attenuation coefficient and QE are relevant to the device thickness, bulk sensitivity should be included in the report, derived by $S_{bulk}=S/d$ (d is the thickness of the material). This parameter is especially important for thin-film X-ray detectors. 
\end{itemize}

The temporal response is another important parameter for X-ray detectors, which defines the response speed (or bandwidth). Though fast-response detectors have been regarded as useful for applications in pulsed X-ray detection, \cite{chen2021fast,zhou2021pulsed} our work demonstrates that a slow detector is also effectively applicable by measuring its stable peak-valley currents. Nevertheless, fast-response detectors are still required in general, especially for the safety in medical and healthcare uses (e.g., FLASH radiotherapy) \cite{bazalova2025balancing}. The response time of Ga$_2$O$_3$-based X-ray detectors has been calibrated with 10$\%$-90$\%$ (preferred in engineering), or more scientifically $1/e$ (e.g., this work) time constants. Both are acceptable in our opinion. To accurately determine the response speed, a sufficient time is suggested after the X-ray beam is turned ON/OFF while monitoring the photocurrent rise/decay, ensuring the photocurrent fully saturates/relaxes (e.g., $\bold{Figs.4a}$ and $\bold{4b}$). Insufficient rise or decay will not give a correct decay time constant,\cite{lu2019x,yu2025high} even for evaluating the fast component (e.g., $\bold{Figs.4d}$ and $\bold{4e}$).

We have also noticed that some self-defined device characteristics start to spread in the Ga$_2$O$_3$-based X-ray detector community. A representative example is the responsivity defined as, $R=\frac{I_{tot}-I_{dark}}{D_{air}}$, often followed by a noise-equivalent dose rate, $NED=D\frac{\sqrt{2eI_{dark}}}{I_{tot}-I_{dark}}$. \cite{lu2019x,zhou2021pulsed,prasad2023ga2o3} The definitions are likely to imitate the photoresponsivity and specific detectivity used in photodetectors. However, the correct responsivity should be the photocurrent generated per power, $I_{ph}/P_{photon}$, which is closely related to QE defined by the number of electrons generated per photon, $n_{e}/n_{photon}$. Moreover, the sensitivity is already linked to the QE for X-ray detectors with Eqs. \ref{eq7} and \ref{eq9} by, 

\begin{equation}
\begin{split}
QE & =n_{e}/n_{photon} \\
   &=\frac{\frac{I_{ph}}{e}}{\Phi A_{o}} \\
  &=\frac{\frac{I_{ph}}{e}}{\frac{\dot{D}_{air}A_{o}}{(\frac{\mu}{\rho})_{air} E_{ph}}} \\
  &=\frac{(\frac{\mu}{\rho})_{air}E_{ph}S}{e}
\end{split}
\end{equation}

\noindent Therefore, the definition of $R$ with photocurrent and dose rate is inappropriate and redundant. As a result, the following NED defined with dark current and dose rate lacks physical meaning, as $I_{dark} \neq I_{noise}$ (where $I_{noise}$ is the noise current) and the device geometry contribution is not included. For a better understanding of NEP, please refer to a recent report. \cite{sakhatskyi2025characterizing} Overall, the newly defined device parameters should be used with caution when benchmarking the device performances in this emerging research field.

\section{Methods}\label{sec4}

\textbf{Material growth} High-quality $\beta-$Ga$_2$O$_3$:Si thin films, with a film thickness of $\sim 2.03 \mu m$, were grown on sapphire (0001) substrates with a $0.2 ^{\circ}$ miscut using an AIXTRON 3x2 CCS MOCVD system. TMGa, TEOS, and 5N O$_2$ were used as the precursors for gallium, silicon and oxygen, respectively. In-situ laser reflectance via a LayTec EpiTT™ system, allowed real-time monitoring of growth rate. Emissivity-corrected pyrometry provided the surface growth temperature. High-purity N$_2$ was used as the carrier gas for TMGa, TEOS and the MOCVD flow gas. The molar flows of the precursors were controlled using mass flow controllers (MFC). The effective Ga/Si molar flow ratio for the two samples was $4.4 \times 10^{-4}$ and $7.5 \times 10^{-7}$, respectively. 

Prior to the growth, the sapphire substrates were subject to a period of annealing at 720 $^{\circ} C$ for 10 minutes under an oxygen flow rate of 500 sccm at a pressure of 150 mbar. Following the annealing, a 30 nm thick nucleation layer was deposited at 650 $^{\circ} C$ under a VI/III ratio of 6,000 and reactor pressure of 100 mbar to minimise lattice mismatch and ensure preferential growth of the desired $\beta-$Ga$_2$O$_3$:Si thin films. The growth of $\beta-$Ga$_2$O$_3$:Si thin films was performed at 950 $^{\circ} C$ at a reactor pressure of 50 mbar. The flow rates of TMGa and O$_2$ for the growth were 30 sccm and 1000 sccm, respectively, which corresponds to a VI/III molar ratio of $\sim$333.

\noindent \textbf{Material characterizations} XRD measurements of the $\beta-$Ga$_2$O$_3$:Si films were performed using Bruker D8 Discover Diffractometer equipped with a Cu K$\alpha$ X-ray source ($\lambda$ = 1.5406 $\AA$) operating at 40 kV and 40 mA. Diffraction patterns of the films were recorded over a 2$\theta$ range of 10$^{\circ}$-70$^{\circ}$, with a step size of 0.030$^{\circ}$ and a scan time of 0.5 seconds per step. All measurements were conducted at room temperature, and data analysis was carried out using Bruker DIFFRAC.EVA 7.1 software. SIMS profiling was carried out by EAG Laboratories to quantify the C and Si concentrations of the samples, providing depth profiles over a range of 0-2 $\mu m$. Transmittance measurements were performed using a Cary 5000 spectrophotometer scanning over the wavelength range of 200–1000 nm. The optical band gaps of the samples were estimated using Tauc plot analysis, as $\beta-$Ga$_2$O$_3$ exhibits a direct-like optical transition in the absorption spectra.

For the TEM measurements, cross-sectional lamellae were prepared using an FEI Helios G4 DualBeam system. A Pt layer was deposited for protection, followed by coarse milling with 30 kV Ga$^+$ ions at 9.3 nA, and fine polishing at 40 pA. The final lamella thickness was $<$100 nm. Film cross-sections were imaged in an FEI Titan Themis TEM operated at 300 kV using an 8.2 mrad probe convergence semi-angle, 70 pA probe current, and 20 $\mu s$ pixel dwell time. 

\noindent \textbf{Device fabrication} The device was fabricated by standard UV lithography and lift-off methods. Prior to fabrication, the surface of the epitaxial wafers was cleared by standard solvent treatments, followed by oxygen plasma ashing. Ti (20 nm)/Au (100 nm) interdigital electrodes were deposited by e-beam evaporation, followed by rapid thermal annealing at 450 $^{\circ}C$ to improve the device stability. The length and width of the fingers are 1000 $\mu m$ and 50 $\mu m$, with bonding area of 1000 $\times$ 1000 $\mu m^2$ for both electrodes. The total optical active area is 1 $\times$ 3.05 $mm^2$ (full device area including all contact region). After fabrication, the devices were bonded to a printed circuit board (PCB) to facilitate the subsequent device characterizations.

\noindent \textbf{Device characterization} The current-voltage (IV) scan was performed with a Keithley 2636B source-meter unit (SMU). The scan speed between two adjacent data points is set to 25 ms with a voltage interval of 50 mV for all the measurements in this work, unless stated otherwise. 

The X-ray response was measured at Diamond Light Source, UK's synchrotron radiation facility. The X-ray introduced from the storage ring was sent to a monochromator, which generates monochromatic beams with a size of $\sim$ 5$\times$5 $mm^2$. The energy used in our experiment ranges from 8 to 25 keV, crossing the Ga 1s electron binding energy. The device and a Si-based dosemeter were mounted on an automatically controlled translation stage, where they are in the same plane for accurate dose measurements. Each dose value is obtained by averaging 100 measurements. An alignment X-ray camera is placed downstream of the device under test (DUT) to verify accurate alignment with the radiation beam. For the transient and modulated X-ray measurements, a software-controlled shutter with a speed of 20 ms was exploited. The photocurrent was registered by a Keithley 2636B SMU.

\section{Conclusion}\label{sec5}

We have monitored the recent rapid development of  Ga$_2$O$_3$-based X-ray detectors, and examined the existing issues, specifically in the methods of device measurements and data analysis. Instead of merely focusing on improving one or more device characteristics, we designed and performed an indicative research work based on epitaxial $\beta-$Ga$_2$O$_3$:Si planar Schottky detectors, measured with \textit{energy-tunable monochromatic} X-ray beams on a synchrotron beamline. The work enables us to clarify the photocarrier generation process in such direct X-ray detectors rarely explored before: direct absorption near the core electron binding energy and inelastic scattering-dominated absorption at other energies. Furthermore, the carrier transport mechanism was unveiled in the dark and under X-ray radiation with various Si doping concentrations. The dark current is dominated by space charge-limited current (SCLC) with low-doping concentration ($\sim 10^{15} cm^{-3}$) while the relatively high-doping ($\sim 10^{18} cm^{-3}$) ones are governed by Poole-Frenkel current, depending on the carrier traps (material defects) partly or fully filled by the doping carriers. In contrast, the photocurrent under X-ray is contributed by only Poole-Frenkel current and is irrelevant to the doping concentration, due to traps being photo-excited (no longer filled). With \textit{saturation photocurrent} from sufficient $I-t$ measurement, the sensitivity of our devices is properly benchmarked, with a value of  $S_A=1.42 \, \mu C\cdot mGy \cdot cm^{-2}$ (high doping) and $S_B=0.09 \, \mu C\cdot mGy \cdot cm^{-2}$ (low doping) at 12 keV, with a photo-to-dark contrast ratio $>3$ orders of magnitude. 

Most importantly, we have proposed a series of protocols for appropriate measurement methodology and rigorous data analysis to benchmark the sensitivity, one of the key characteristics of X-ray detectors. In parallel, the definitions of several parameters and associated equations are clarified and corrected to enable accurate and consistent reporting of device performance. By systematically addressing commonly overlooked issues, such as response saturation, measurement timing, and parameter interpretation, our initiative establishes a more reliable framework for evaluating Ga$_2$O$_3$-based X-ray detectors.

We envisage that this work will offer clear guidance and a standardized reference for the thriving Ga$_2$O$_3$-based X-ray detector research community, facilitating meaningful comparison between reported results and accelerating the translation of laboratory demonstrations into practical applications. The proposed protocols are broadly applicable and can be readily extended to other wide-bandgap semiconductor X-ray detectors, thereby contributing to improved reproducibility and benchmarking consistency across the field. Ultimately, this study aims to promote more transparent performance assessment and to support the rational design and optimization of next-generation X-ray detection technologies.

\backmatter

\bmhead{Data availability}

The data that support the findings of this study are available from the corresponding authors upon reasonable request.

\bmhead{Supplementary information}

Supplementary $\bold{Fig. S1-9}$,  $\bold{Tables \, S1-2}$ and $\bold{Note \, S1}$. Appendix: `linear attenuation coefficient-Ga$_2$O$_3$'. 

\bmhead{Acknowledgements}

This work is supported by the Royal Society (Grant No. IEC/NSFC/242145), and the National Epitaxy Facility pump-priming project (PP57/2025). D.L and L.L.thank the support from EPSRC under Grant No. EP/T019085/1.

\bmhead{Contributions}

Y.H. conceived the idea, organized the research project, led the data analysis and manuscript writing. A.M. fabricated the device and joined the device testing and manuscript drafting. D.L. led the growth and material measurements. L.L. co-led the project and funding. O.F. and K.S. offered the X-ray sources and dose measurements. C. L joined material growth and measurements. J.E. assisted in electrical measurements. S.R.  measured XPS. T.C., J.H., Z.A., A.B., and R. D-B contributed to the TEM and EDS characterizations. All authors participated in the manuscript writing.

\section*{Declarations}
The authors declare no competing interests.



\end{document}